# IEEE Copyright Notice

# Tactical Network Modeller Simulation Tool

## Combined discrete event and network back-ends

*Stuart Marsden*

Dept. of Military Technology
National Defence University
Helsinki, Finland
stuartmarsden@finmars.co.uk

*Jouko Vankka*

Professor of Military Technology
National Defence University
Helsinki, Finland
jouko.vankka@mil.fi

*Abstract*—This paper discusses the implementation of a tactical network simulation tool. The tool is called Tactical Network Modeller (TNM). TNM uses some novel techniques to simplify the building of the network model using graph theory constrained by a hierarchical tree which reflects the organisation structure. TNM allows models to be constructed using an Application Programming Interface (API) or a node based User Interface (UI). When the model is constructed, different simulation back-ends can be applied to it. A discrete event simulation and a network emulation back-end are implemented building on top of open source tools.

TNM is simple to create models for non technical users. The model can be used to analyse information flows. The same model can be used for a full network emulation. This allows real software and protocols to be tested in a realistic simulated environment. The flexibility of the software allows its use from engineering up to campaign planning.

## I. INTRODUCTION

Simulation and Modelling are important components in the planning and deployment of military systems. In the Command, Control, Communications, Computers, and Intelligence (C4I) space the complexity of heterogeneous systems that are deployed can have unpredictable outcomes. Planners use their experience to build C4 architectures but do not have an easy way to validate them. Simulation can be used to improve doctrine and processes generally and for specific operations adapt the architecture to fit the environment.

Modelling is often used during the design and implementation of new systems. It can be used as a testbed to provide a detailed simulation during system evaluation[1]. Where simulation is used to look at scenarios a generic approach is often used with networks generated by algorithms such as "Small-World" which is not representative of military networks[2].

Neither technique are commonly used as a tool for pre-deployment nor on-deployment network planning at the tactical level. Modelling and Simulation is a specialist area and requires personnel trained in specific tools and techniques.

This paper presents a novel custom designed tool called Tactical Network Modeller to readily allow the modelling of a tactical network and its interconnections. This model can be built visually with very basic training. This model can be used to simulate the environment with different back-ends. This enables generic scenario investigation allowing planning decisions to be made. A different back-end allows the full network stack to be emulated so that actual software and protocols can be investigated.

## II. BACKGROUND

There are three major techniques for researching networking concepts[3]. Live network testing especially in a military environment is often prohibitively expensive and is not covered in this paper. This leaves simulation and emulation:

### A. Simulation

Time slicing techniques are simplistic and inefficient so discrete event simulation is preferred[4]. This processes only when a significant event takes place within the simulation model. The tracked simulation time is unit-less and can be arbitrarily chosen as the simulation dictates. This approach can run much faster than real-time allowing fast scenario analysis.

### B. Network Emulation

A discrete event tool such as NS-3 can be used for network emulation but with limitations[5]. A Network Simulation seeks to emulate the network stack with as much fidelity as possible. It will run in real-time allowing standard applications and protocols to use the emulated network seamlessly. It should present a standard network layer but should allow that network to be built and constrained to a specific model. This should include network effects such as delay, jitter and packet loss.

## III. RELATED WORK

Discrete event simulation tools have been widely used in order to understand how new technologies and network protocols may work. An example of this would be OverSim which can be used to investigate peer-to-peer networks and is built with the OMNeT++ framework[6]. In contrast are tools such as the stochastic discrete event simulator proposed in [7] which looks at structures rather than specific technology. That tool allows scenarios for effects based planning but requires knowledge of the XML-model to create structure and rules.

Another network emulation approach uses NetEm features in the Linux operating system(OS) kernel that applies network effects. This allow properties like packet loss, rate control, duplication and reordering to be applied [8]. This can be combined with network namespaces for lightweight emulation as is done in the Common Open Research Emulator (CORE) tool [9].

This prior work has produced specific tools which can only be used for a subset of the types of simulation needed. A model may have to be built in several tools and kept synchronised. TNM builds upon and unifies these approaches.

## IV. Approach

The goal is to produce a usable modelling application which would allow investigation of scenarios and also allow technical protocol development and testing. The TNM tool provides a novel combination of simulation back ends allowing the resolution of the model to be tailored to the level of the hierarchy. In [10] a hierarchy of modelling and simulation was presented going from engineering through engagement and mission to campaign. TNM can have utility at all levels of that hierarchy. Emulation can allow real-time and detailed protocol development, discrete event simulation can model objects as sub-systems or as entire military formations.

The Objects used to build the model rely on object orientation within the implementation. This is also reflected in the structure of the model. In [11] a way of viewing problems using colours for archetypes is presented. The model has objects that reflect the Thing, Role, MomentInterval and Description archetypes. This approach allows the UI and API to present a coherent structure which can scale easily.

The produced model is a directed graph and shows the interconnections between simulated elements. Graph construction is constrained in order to facilitate re-use and to keep the graph structure logical. In order to do this, a hierarchy is imposed upon the graph. This hierarchy means that each node must have a parent and can have one or more children. The model also provides a graph of the network interconnections. This graph is then used to control message passing or link emulation between nodes. The modelled network can have multiple simulation back-ends attached. For example a model may include several discrete event back-ends for different concepts e.g. Voice, Data or Information flow. Within this back-end the detail of each node's resources and tasks are defined. The simulation can be run and the results logged for analysis.

TNM provides an API which allows a model to be built within a Python program. A UI based tool is also provided which allows the model to be built visually. The two approaches can be mixed as both the API and the UI can save and load a common XML file which describes the model.

## V. Hierarchical modelling

As military organisations tends to be hierarchical the tool provides a means to build a hierarchy quickly. This allows a sub-unit to be built and then duplicated. So for example an infantry section can be modelled and then duplicated so that there are 3 within a platoon which can then be duplicated so that there are 3 in a company and so on.

When duplicating a hierarchy the internal connections are duplicated entirely whereas the external connections are set for that sub-unit. In order to ensure that this process is manageable and returns a sensible connection graph, limitations are placed on the types of connections. The hierarchical connections are a tool for the modeller rather than a driver for the simulation as they allow structure to be given to the model but do not directly affect how the network simulation will work.

### A. Node Based

Three types of nodes have slightly different rules applied, outlined below. They were developed with an object oriented language and as such the base class is the Network Object. The Composite and Area Network Objects are specialisations. All the nodes therefore inherit the behaviour of Network Object.

At the top of the hierarchy is always the Model object which acts as the default parent for new Objects. It also provides helper functions for creating the graphs for connection and hierarchy and finding named objects within the model.

#### 1) Network Object

The base class provides the standard functionality. A Network Object allows the setting of a name which must be unique within its siblings. The node provides an interface concept which allows the building of network connections. The Network Object can have an arbitrary number of interfaces which must be uniquely named. An interface can have a connection (which will become an edge in the model graph) to any other Object interface. An interface can have multiple connections and by default are bi-directional. Connections are restricted to either siblings or the immediate parent. Inter-connection between Objects with different parents is prohibited.

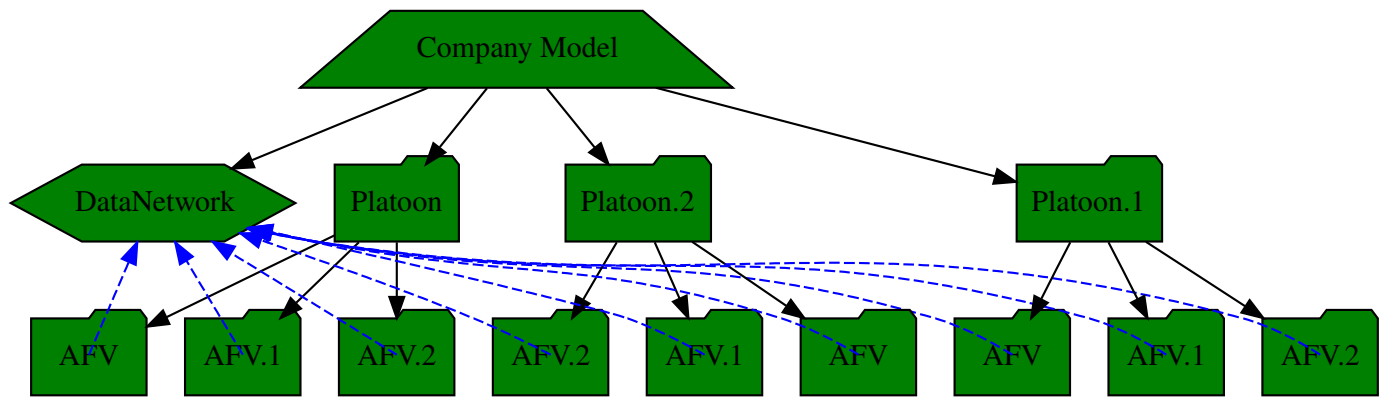

Fig. 1. Example model graph with AFV detail rolled up

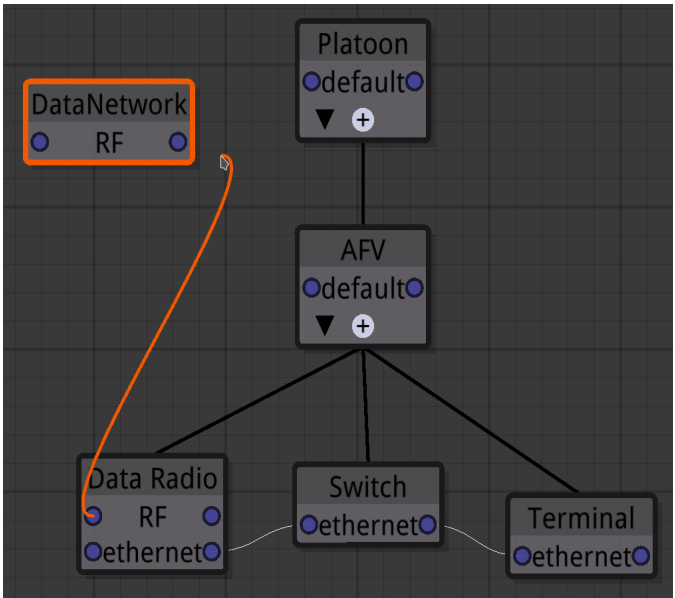

Fig. 2. Portion of a model being built in UI

### 2) Composite Object

The Composite Object has all the functionality of the Network Object but in addition it can act as a parent for any other Object including other Composite Objects. A Composite object can act as an Interface for its children to the wider network. This simplifies outward connections, as an example a platoon Composite Object could have a company network interface. The children that will be attached to the company network will have a connection to the parent representing it. The parent can then be connected once to the Object representing this interface. To change the onward connection of this interface requires only the parent to be changed with no knowledge needed of the internal structure of that composite. This approach allows components of the model to be considered as black boxes and present a consistent interface to the rest of the model.

### 3) Area Network Object

The Area Network Object exists as an alternative to using the parent as the interface. It is a specialisation of the Network Object which does allow connections outside of the parent-child hierarchy. This can be used to represent the concept of a network such as a VHF radio network. In a model there may be a company network which is used at section, platoon and company level. An Area Network would be placed as a child to the company Composite Object. Network connections can then be made within any level of hierarchy of the company.

When a Composite Object is copied containing an Area Network Object a new version of the Area Network Object is made and the internal connections within the duplicated hierarchy are also duplicated.

### B. User Interface Modelling

The described capabilities can be used within the user interface to visually build models. The hierarchical structure is constructed by dragging a line from the parent object to the child. The network connections are dragged between the interfaces. Fig. 1 shows a small portion of a model being built. The hierarchy can be seen with the black wires while the network connections have white wires. A new connection is in the process of being made (orange wire).

Once a portion has been modelled the detail can be hidden by rolling up each child and just displaying the network. For example the Armoured Fighting Vehicle (AFV) could now be rolled up as the model is complete. The AFV can then be selected and duplicated. This will make a new AFV which is still a child of platoon and has the same child objects and internal connections. The specialised connection to the Area Network Object will be maintained so the child Data Radio Object will have its own connection to the Data Network. The child names will be maintained as they are unique within their siblings but the new Composite Object cannot be named AFV and will be automatically changed to AFV.1, this can be changed as long as it is unique within its siblings.

### C. Building Model using API

It is possible to construct the model using the Python API. This allows the model to be constructed with defined objects and connections. An example code snippet that defines the same model as Fig. 1 would look like:

```
# Model is the root object
examplemodel = Model('Company Model')

# Top level objects are parented directly to the model
# First argument is always the parent
data_network = AreaNetwork(examplemodel, name='DataNetwork')
platoon = CompositeObject(examplemodel, name='Platoon')

# AFV is a child of platoon
afv = CompositeObject(platoon, name='AFV')

# The next three are children of AFV
data_radio = NetworkObject(afv, name='Data Radio')
router = NetworkObject(afv, name='Router')
terminal = NetworkObject(afv, name='Terminal')

# Interfaces are bi-directional
# so can be created from either end
router.add_interface(data_radio)
router.add_interface(terminal)
data_radio.add_interface(data_network)
```

The duplication of Objects is possible in the same way as the UI. This should be done in the code when all the detail of the object and its children has been defined. That detail will then be copied to the new object. To define a new AFV based on the already modelled object:

```
afv2 = afv.copy()
```

The graph produced by that code can be visualised using a create_diagram function and is reproduced at Fig. 2. This could then be scaled by copying the platoon object twice to model a Company.

## VI. NETWORK CONNECTIONS GRAPH

Network connections are distinct to the hierarchical structure. Network connections can be made between all object types but is restricted according to some set rules. A child object can only connect to its siblings, parent, or if it is a composite object to its own children. The Area Network Object is an exception and allows connections regardless of the relative positions in the hierarchy.

Each object has a single interface point by default but more can be added as required. Each interface point can have multiple connections if needed. All connections are bi-directional.

The Network Objects form the vertices in a directed graph as each network connection forms two directed edges (also refereed to as arcs) to represent the duplex connection[12]. This could allow for the modelling of uni-directional links but this is not implemented at this level. As each vertex has the concept or multiple interfaces this would allow parallel arcs between vertices. The produced graph could therefore be a directed multigraph but as loops are not allowed they are not pseudographs. Due to each connection producing two edges the graph is strongly connected.

The software provides the functionality to determine all paths between two vertices in the graph. This is found using a Depth-First Search algorithm without weighting the arcs. This functionality is used by the discrete event simulation back-end as an analogue to network routing such as Open-Shortest Path First (OSPF).

## VII. Simulation back-ends

TNM allows two types of simulation back-end to be used with any model: discrete event simulation and network emulation. For each model it may have multiple back-ends of either type associated with it. This allows simulations with different configurations to be specified and run.

### A. Discrete event

The discrete event back-end builds on SimPy [13] which is an open source simulation framework for the Python programming language. The simulation is specified using 3 different components. These components are attached to an Object and will be saved and copied when building the model. This makes it easy to specify a sub-set of the entire model with the simulation components and then duplicate them to build a larger structure. The 3 Object types to build a discrete event simulation are:

#### 1) Resources

A resource can be used to represent any contended resource such as a human who only has the capacity to process one task at once or a computer that can handle more concurrency but still has limitations or a network. A resource could be applied to an Area Network Object and given a capacity which indicates how many concurrent users it can have. A delay can also be specified. As the simulation is run it will ensure that access to resources is granted based on the specified capacity. All other waiting processes will be blocked until the resource is available.

#### 2) Tasks

A task is an event triggered at a set time in the simulation and optionally is repeated a given number of times. The time between repeats can have a random Gaussian distribution applied to vary a set delay. Implemented tasks include:

*a) Message.* This can be routed based on the network which is determined by the shortest path through the graph. To simulate a packet based network each resource along the path is acquired in turn. Once acquired the resource delay is applied and that resource is then released. This is then repeated for the next network hop until the destination is reached. A non routed message is also possible which only implements delays and capacity at the receiving node not any intermediates.

*b) Acknowledgements.* Either type of message can request an acknowledgement. The remote service should then send a message back.

#### 3) Services

Services can also be assigned to objects. These services represent processes that can be queried. An example service would be waiting for messages to arrive and then sending the acknowledgement. The service has a message queue and is triggered when an external process places something in this queue. The service can choose to apply a delay for each message which will limit the capacity. Due to the simulation mechanism a service with no delay has infinite capacity and can deal with requests immediately.

### B. Network Emulation

The alternative method of using the model is to run a network emulation. This uses the network objects and connections to build a full network emulation. The emulation uses CORE[9] to construct and run the emulation. CORE provides a Python API which has eased the integration with TNM as it is written in the same language.

CORE uses kernel virtualisation to provide each emulated network object with its own network interface. These nodes are then interconnected with bridging and packet manipulation. Real network hardware can also be integrated in to the network.

Each node can now run any software designed for the Linux or BSD OS. This allows server software to be run such as web, email or database servers. Client software can also be run whether being network loading tools or a full UI application. One method which makes the deployment of services to each virtual network object easier is using application containerisation using Docker[14]. The use of Docker allows the same software to be deployed in a known state to multiple nodes in a virtual network. Every run of the emulated environment is then easily repeatable with no contamination from other runs. This makes Docker an ideal tool for research environments[15].

## VIII. Result Capture

TNM provides the capability to capture the results and analyse them to make meaningful conclusions. The approach taken by TNM varies depending on the selected back-end.

### A. Logger

TNM has a logging ability which records data in to a Pandas dataframe [16]. This is saved in the Hierarchical Data Format (HDF). The logging is flexible and allows textual and scalar values to be recorded. Each log entry is stamped with the time as given by the back-end. Pandas is a python data manipulation library so some standard analysis can be built in to TNM and each back-end can implement their own data transformations.

## B. Network Tools

The logger can provide valuable data for the network emulation back-end if it is configured but other approaches can give more detail. As the emulation represents a full network stack it is possible to use all the tools that are available on a real network. This allows the use of tools such as Wireshark or tcpdump to capture actual data traffic either in real time or recorded for later analysis.

## C. Analysis tools

With the data captured there are many possible tools to conduct analysis. The chosen ones will depend on the objectives of the simulation and the familiarity of the user.

As the primary log is stored in HDF it is easy to import in to data analysis tools. They can be reimported back in to a Pandas dataframe and manipulated. This can be done in python and plots produced using plotting tools. Some examples can be seen in the next section.

## IX. EXAMPLE SIMULATION

To show some of the capabilities and illustrate a possible analysis an example model was built and a scenario defined using the discrete event simulation back-end. Whilst the model produced was simplistic with few tasks it was possible to find bottle necks and make assessments about network capacity.

## A. Set up Model

The model was defined using the python API but could equally have been made using the UI. The AFV defined in Fig. 1 was used and built up with 3 AFVs per platoon and 3 platoons per company. In addition each company was given a company headquarters (HQ) which was connected to the data network for the company but also to a separate data network that forms part of the BattleGroup (BG) level. The BG was made by duplicating 3 companies and defining a BG HQ. The overall network is shown at Fig. 3. This shows network connections emerging from each company and going to the BGDataNetwork object. They originate from the 2nd radio in each Coy HQ.

The example is at BG level but by using the same process of parenting and duplication a model can be built up to any level readily. The duplicated composite objects such as platoons can be altered after duplication so variations in structure can be accounted for. TNM allows saving to an XML file and reloading thus it would be possible to define common sub-units and store them for later recombination to make the structure required.

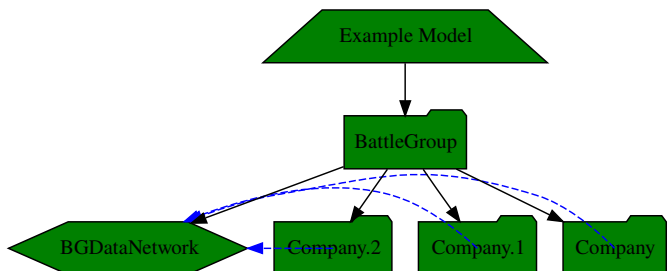

Fig. 3. Overview of the example simulation model.

## B. Simulate

A single simulation back-end was defined using the SimPY discrete event simulation. Each data network was given a resource that defined the capacity and delay. Each router also was given a resource with a set delay and capacity.

Terminals were given a service which would listen for messages and send an acknowledgement if requested. Each AFV terminal had a task to send a message to its corresponding company HQ at a set time repeated to simulate standard reports and returns. A regular message was sent between AFVs of different companies which simulates cross boundary position reporting. Due to the structure of the network the cross boundary messages have a long network path to the recipient.

The simulation was set to run for a simulated time period of 10 hours in order to have several repeats of each message type. Due to the discrete event simulation approach the actual run time was a few seconds. All messages were routed and would therefore traverse the network in the shortest path but have to acquire each resources in turn before the message could be sent. This creates a message based packet switched network.

In the scenario the reports and returns are non-urgent and sending can be delayed. The reports and returns request an acknowledgement message. The position reporting message does not request an acknowledgement but for the scenario has a requirement of delivery within 10 seconds and is time critical.

## C. Results

The simulation was run and produced a log dataframe. The log contained 49,800 records including resource usage and every hop of every sent message during the simulation run. In this time 2041 messages were sent. The simulation was run several times to explore the scenario and Pandas was used to extract the message sending time for the position reporting messages only. The resulting dataframe was exported and used to produce charts in a spreadsheet.

## D. Analysis

The analysis covered the requirement to ensure that position reports reach their destination in time. This involved several runs of the simulation with slightly altered parameters. The simulation was first run without the reports and return messages and the company and BG data network capacity set to only 1 message at once. The results are in blue in Fig. 4 and show that the delivery is under the 10 second requirement but with little margin. As the only place these messages contend with each other is on the BG data network its capacity was increased to 2 concurrent messages. The result is in orange and shows a reduction of the delivery time to an acceptable 6.4 seconds.

The reports were then turned on and they clearly contended with the position reporting as can be seen by the spikes in the yellow section of the chart. They spike at 21.7 seconds thus not meeting the requirement. As the reports are only sent within each company network one option is to increase the network capacity. It is found, as seen in the green section, that the capacity would have to be increased to 4 which may not be possible. As reports are not time critical they can be adjusted. A random 30 second delay was applied to reports messages and this was sufficient to reduce the maximum sending time even with the company data network capacity remaining at 1 (red section).

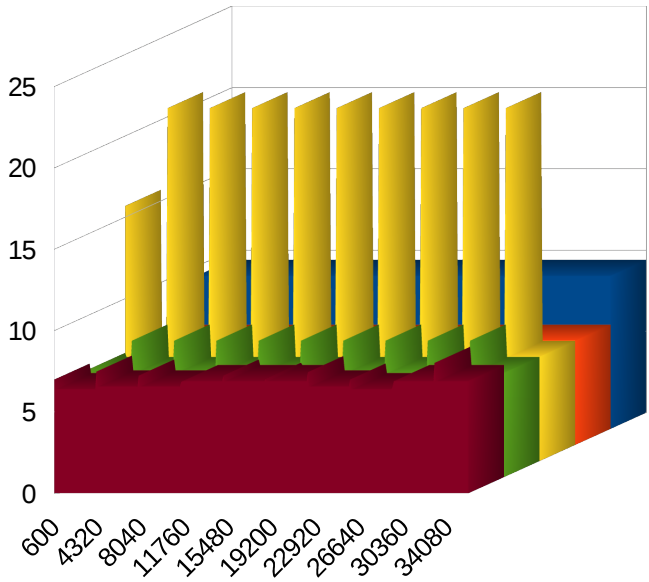

Fig. 4. Chart of position report delivery time over simulation run

The analysis of this simple network lead to real conclusions. At times capacity needs to be increased but it is also possible to meet the requirements with slight changes to the procedures.

### E. Other Scenarios

The tool can be used to analyse many different scenarios. This can be message sending or more abstract information flow. The tool could be used to explore the timeliness of different information passing approaches such as voice versus data communications. The optimisation of voice and data networks can be performed by running the same tasks over different topologies.

The network emulation tool can be used to validate the use of new protocols on restricted networks such as Voice over Internet Protocol (VoIP) or Enterprise Service Bus. It could even be used to represent an entire deployment's software infrastructure on a single powerful workstation.

## X. CONCLUSIONS

Whilst there are many simulation packages available they do not specifically address the needs of the tactical user. The ability to build models in a simple and intuitive way means the task can be undertaken by a staff officer rather than a simulation expert. The discrete event simulation allows this model to be used for a broad spectrum of military scenarios without the need of detailed network knowledge.

The same model can be used with the actual software and services that would be deployed. This allows detailed exploration of the proposed networks ability to meet the demands placed on it. This facilitates testing of new software and network protocols across a valid simulated network.

## XI. FUTURE WORK

TNM could be extended to offer more choice of simulation back-ends such as OPNET, OMNeT++ or ns-3. This could provide a spectrum of utility between discrete event simulation and full network emulation.

The SimPY back-end could allow the model designer to script additional functionality to the service objects. This could be done using the node UI to build a simple programming UI similar to that described as Flow Based Programming[17].